\documentclass[a4paper,conference]{IEEEtran}

\IEEEsettopmargin{t}{30mm}
\IEEEquantizetextheight{c}
\IEEEsettextwidth{14mm}{14mm}
\IEEEsetsidemargin{c}{0mm}

\usepackage[utf8]{inputenc}
\usepackage[T1]{fontenc}
\usepackage{ifthen}
\usepackage[cmex10]{amsmath} 
\usepackage[url,hyperrefblack,notheorems,IEEEtran]{research17} 
\usepackage{balance} 
\usepackage{amsmath,amssymb,amsfonts,amsbsy}
\usepackage{ifthen} 
\newboolean{short_version}
\setboolean{short_version}{false} 
\usepackage{enumitem}
\usepackage{mathtools}
\usepackage[lined,boxed,commentsnumbered,linesnumbered,ruled]{algorithm2e}
\usepackage{comment}
\setlist[description]{leftmargin=\parindent,labelindent=\parindent}

\newtheorem{theorem}{\mytheoremname}
\newtheorem{lemma}{\mylemmaname}
\newtheorem{corollary}{\mycorollaryname}
\newtheorem{conjecture}{\myconjecturename}
\newtheorem{proposition}{\mypropositionname}

\newtheorem{definition}{\mydefinitionname}
\newtheorem{remark}{\myremarkname}
\newtheorem{example}{\myexamplename}
\newcommand{\Hwt}[1]{\wH\left(#1\right)} 
\newcommand{\vol}[1]{\operatorname{vol}\left(#1\right)} 
\newcommand{\dual}[1]{{#1}^\perp} 
\newcommand{\ConstrA}[1]{\Lambda_\textnormal{A}(#1)} 

\newcommand*{\Scale}[2][4]{\scalebox{#1}{\ensuremath{#2}}} 
\makeatletter
\renewcommand*\env@matrix[1][*\c@MaxMatrixCols c]{%
  \hskip -\arraycolsep
  \let\@ifnextchar\new@ifnextchar
  \array{#1}}
\makeatother
\usepackage{tikz}
\usetikzlibrary{calc,shapes, patterns,decorations.text, decorations.pathreplacing}
\usepackage{ctable} 
\usepackage{afterpage} 
\usepackage{lscape} 
\usepackage{pdflscape} 
\usepackage{rotating} 
\usepackage{pbox} 

\usepackage{todonotes}

\begin{document}

\title{The Secrecy Gain of Formally Unimodular Lattices\\ on the Gaussian Wiretap Channel}

\author{%
 \IEEEauthorblockN{Maiara F.~Bollauf, Hsuan-Yin Lin, and {\O}yvind Ytrehus}
 \IEEEauthorblockA{Simula UiB, N--5008 Bergen, Norway\\             
             Emails: \{maiara, lin, oyvindy\}@simula.no}
}



\maketitle

\begin{abstract}
  We consider lattice coding for the Gaussian wiretap channel, where the challenge is to ensure reliable communication between two authorized parties while preventing an eavesdropper from learning the transmitted messages. Recently, a measure called the \emph{secrecy function} of a lattice coding 
  scheme was proposed as a design criterion to characterize the eavesdropper's probability of correct decision. In this paper, the family of \emph{formally unimodular lattices} is presented and shown to possess the same secrecy function behavior as unimodular and isodual lattices. Based on Construction A, we provide a universal approach to determine the \emph{secrecy gain}, i.e., the maximum value of the secrecy function, for formally unimodular lattices obtained from formally self-dual codes. Furthermore, we show that formally unimodular lattices can achieve higher secrecy gain than the best-known unimodular lattices from the literature.
\end{abstract}

\section{Introduction}

In recent years, \emph{physical layer security}  based on 
information theory has attracted a great deal of attention for secure applications in  wireless communications in 5G and beyond (see~\cite{WuKhisti-etal18_1} and references therein). This line of research has evolved from 
the classical \emph{wiretap channel (WTC) model} introduced by Aaron Wyner in his landmark work~\cite{Wyner75_1}, which showed that reliable and secure communication can be achieved simultaneously without the need of an additional cryptographic layer on top of the communication protocol.

Since then, substantial research efforts have been devoted to developing practical codes for reliable and secure data transmission over WTCs. Among the potential candidates are \emph{lattices}, where in~\cite{BelfioreOggier10_1,OggierSoleBelfiore16_1} it was shown that a lattice-based coset encoding approach can provide secure and reliable communication on the Gaussian WTC. 
In particular, it was shown that for Gaussian WTC, the so-called~\emph{secrecy function} expressed in terms of the \emph{theta series} of a lattice (see the precise definition in Section~\ref{sec:secrecy-function_lattice}) can be considered as a quality criterion of good wiretap lattices codes: to minimize the eavesdropper's probability of correct decision, one needs to maximize the secrecy function, and the corresponding maximum value is referred to as \emph{(strong) secrecy gain}.

Belfiore and Sol{\'{e}} \cite{BelfioreSole10_1} studied \emph{unimodular} lattices and showed that their secrecy functions have a symmetry point. The value of the secrecy function at this point is called the \emph{weak secrecy gain}. Based on this, the authors of \cite{BelfioreSole10_1} conjectured that for unimodular lattices, the secrecy gain is achieved at the symmetry point of its secrecy function. I.e., the secrecy gain of a unimodular lattice is equivalent to its weak secrecy gain. Finding good unimodular lattices that attain large secrecy gain is of practical importance. In~\cite{Ernvall-Hytonen12_1}, a novel technique was proposed to verify or disprove the Belfiore and Sol{\'{e}} conjecture for a given unimodular lattice. Using this method, the conjecture is validated for all known even extremal unimodular lattices in dimensions less than $80$. In another work~\cite{LinOggier13_1}, the authors use a similar method as~\cite{Ernvall-Hytonen12_1} to classify the best unimodular lattices in dimensions from dimensions $8$ to $23$. For unimodular lattices obtained by Construction~A from binary doubly even self-dual codes up to dimensions $40$, their secrecy gains are also shown to be achieved at their symmetry points ~\cite{Pinchak13_1}.

This work first introduces a new and wider family of lattices, referred to as \emph{formally unimodular lattices}, that consists of lattices having the same theta series as their dual. We then prove that formally unimodular lattices have the same symmetry point as unimodular or isodual lattices. Similar to the feature of formally self-dual codes defined in coding theory, it is expected that such a broader class of lattices can achieve higher secrecy gain than the unimodular lattices. We pursue this expectation via Construction A lattices obtained from formally self-dual codes and give a universal approach to determine their secrecy gain. For formally unimodular lattices obtained by Construction A from even formally self-dual codes, we also provide a sufficient condition to verify Belfiore and Sol{\'{e}}'s conjecture on the secrecy gain. (A code is called \emph{even} if all of its codewords have even weight, otherwise the code is \emph{odd}.)

Furthermore, we present numerical evidence supporting the conjecture of secrecy gain also for  Construction A lattices obtained from \emph{odd} formally self-dual codes. 
For dimensions up to $70$, we note that formally unimodular lattices have better secrecy gain than the best known unimodular lattices described in the literature, e.g.,~\cite{LinOggier13_1}. We also observe that large minimum Hamming distance and low number of low-weight words in the formally self-dual code corresponds to high secrecy gain of the corresponding formally unimodular Construction A lattice. \ifthenelse{\boolean{short_version}}{Due to page limitations, 
some proofs are omitted and can be found in the extended version~\cite{BollaufLinYtrehus21_1sub}.}{}

\section{Definitions and Preliminaries}
\label{sec:definitions-preliminaries}

\subsection{Notation}
\label{sec:notation}

We denote by $\Integers$, $\Rationals$, and $\Reals$ the set of integers, rationals, and reals, respectively.\ifthenelse{\boolean{short_version}}{}{~Moreover, $\Integers_{\geq 0}$ denote the nonnegative integers, and $[a:b]\eqdef\{a,a+1,\ldots,b\}$ for $a,b\in \Integers$, $a \leq b$.} Vectors are boldfaced, e.g., $\vect{x}$. Matrices and sets are represented by capital sans serif letters and calligraphic uppercase letters, respectively, e.g., $\mat{X}$ and $\set{X}$.\ifthenelse{\boolean{short_version}}{}{~$\mat{0}$ represents an all-zero matrix.} We use the customary code parameters $[n,k]$ or $[n,k,d]$ to denote a linear code $\code{C}$ of length $n$, dimension $k$, and minimum Hamming distance $d$. 
Throughout this paper, we will focus on binary codes only.

\subsection{On Codes and Lattices}
\label{sec:codes-lattices}

Let $\code{C}$ be an $[n,k]$ code and $\dual{\code{C}}\eqdef\{\vect{u}\colon\inner{\vect{u}}{\vect{v}}=0,\forall\,\vect{v}\in\code{C}\}$. The \emph{weight enumerator} of a code $\code{C}$ is given by
\begin{IEEEeqnarray*}{c}
  W_\code{C}(x,y)=\sum_{w=0}^n A_w x^{n-w}y^w,
\end{IEEEeqnarray*}
where $A_w\eqdef\{\vect{c}\in\code{C}\colon\Hwt{\vect{c}}=w\}$. The relation between $W_\code{C}(x,y)$ and $W_{\dual{\code{C}}}(x,y)$ is characterized by the well-known \emph{MacWilliams identity} (see, e.g.,~\cite[Th.~1, Ch.~5]{MacWilliamsSloane77_1}):
\begin{IEEEeqnarray}{c}
  W_{\code{C}}(x,y)=\frac{1}{2^{n-k}}W_{\dual{\code{C}}}(x+y,x-y).
  \label{eq:MacWilliams-identity}
\end{IEEEeqnarray}
We have the following families of codes.
\begin{definition}[Self-dual, isodual, formally self-dual codes]
  \begin{itemize}
  \item A code $\code{C}$ is said to be \emph{self-dual} if $\code{C}=\dual{\code{C}}$.
  \item If there is a permutation $\pi$  of coordinates such that $\code{C}=\pi(\dual{\code{C}})$, $\code{C}$ is called \emph{isodual}.
  \item A code $\code{C}$ is \emph{formally self-dual} if $\code{C}$ and $\dual{\code{C}}$ have the same weight enumerator, i.e., $W_\code{C}(x,y)=W_{\dual{\code{C}}}(x,y)$.
  \end{itemize}
\end{definition}

Clearly, a self-dual code is also isodual, and an isodual code is formally self-dual. Any code in these classes is an $[n,\nicefrac{n}{2}]$ code and, by (\ref{eq:MacWilliams-identity}), its weight enumerator $W_\code{C}(x,y)$ satisfies ~\cite[eq.~(7), p.~599]{MacWilliamsSloane77_1}
\begin{IEEEeqnarray}{rCl}
  W_\code{C}(x,y)=W_{\code{C}}\left(\frac{x+y}{\sqrt{2}},\frac{x-y}{\sqrt{2}}\right).  
  \IEEEeqnarraynumspace\label{eq:FSD_MacWilliams-identity}
\end{IEEEeqnarray}

%


A (full rank) \emph{lattice} $\Lambda$ is a discrete additive subgroup of $\mathbb{R}^{n}$, which is generated as
  $\Lambda=\{\vect{\lambda}=\vect{u}\mat{G}_{n\times n}\colon{\bm u}=(u_1,\ldots,u_n)\in\mathbb{Z}^n\}$,
where the $n$ rows of $\mat{G}$ form a lattice basis. 
The \emph{volume} of $\Lambda$ is $\vol{\Lambda} = \ecard{\det(\mat{G})}$.



If a lattice $\Lambda$ have generator matrix $\mat{G}$, then the lattice $\Lambda^\star\subset\mathbb{R}^n$ generated by  $\trans{\bigl(\inv{\mat{G}}\bigr)}$ is called the \emph{dual lattice} of $\Lambda$.

\begin{remark}
  \label{rm:volume}
  $\vol{\Lambda^\star}=\inv{\vol{\Lambda}}$.
\end{remark}

For lattices, the analogue of the weight enumerator of a code is the \emph{theta series}. 
\begin{definition}[Theta series]
  \label{def:theta-series}
  Let $\Lambda \subset \Reals^n$ be a lattice, its \emph{theta series} is given by
  \begin{IEEEeqnarray*}{c}
    \Theta_\Lambda(z) = \sum_{{\bm \lambda} \in \Lambda} q^{\norm{\vect{\lambda}}^2},
  \end{IEEEeqnarray*}
  where $q\eqdef e^{i\pi z}$ and $\Im{z} > 0$. 
\end{definition}

Analogously, the spirit of the MacWilliams identity can be captured by the \emph{Jacobi's formula}~\cite[eq.~(19), Ch.~4]{ConwaySloane99_1}
\begin{IEEEeqnarray}{c}
  \Theta_{\Lambda}(z)=\vol{\Lambda^\star}\Bigl(\frac{i}{z}\Bigr)^{\frac{n}{2}}\Theta_{\Lambda^\star}\Bigl(-\frac{1}{z}\Bigr).
  \label{eq:Jacobi-formula}
\end{IEEEeqnarray}

Note that sometimes the theta series of a lattice can be expressed in terms of the \emph{Jacobi theta functions} defined as follows.
\begin{IEEEeqnarray*}{rCl}
  \vartheta_2(z)& \eqdef &\sum_{m\in\Integers} q^{\bigl(m+\frac{1}{2}\bigr)^2}=\Theta_{\mathbb{Z} + \frac{1}{2}}(z),
  \nonumber\\
  \vartheta_3(z)& \eqdef &\sum_{m \in \mathbb{Z}} q^{m^2}=\Theta_{\mathbb{Z}}(z), ~ ~ \vartheta_4(z) \eqdef \sum_{m\in \mathbb{Z}} (-q)^{m^2}.
\end{IEEEeqnarray*}


In lattice theory, we have similar concepts to self-dual and isodual dual codes. Here, we also introduce  \emph{formally unimodular} lattices.
\begin{definition}[Unimodular, isodual, formally unimodular lattices]
\label{def:uni-iso-fum}
  A lattice $\Lambda \subset\Reals^n$ is said to be \emph{integral} if the inner product of any two lattice vectors is an integer. 
  \begin{itemize}
  \item An integral lattice such that $\Lambda = \Lambda^\star$ is called \emph{unimodular} lattice.
  \item A lattice $\Lambda$ is called \emph{isodual} if 
  it can be obtained from its dual $\Lambda^\star$ by (possibly) a rotation or reflection.
  \item A lattice $\Lambda$ is \emph{formally unimodular} if it has the same theta series as its dual, i.e., $\Theta_{\Lambda}(z)=\Theta_{\Lambda^\star}(z)$.
  \end{itemize}  
\end{definition}

\begin{remark}
  The relations among unimodular, isodual, and formally unimodular lattices are given as follows.
  \begin{IEEEeqnarray*}{c}
  \bigl\{ \Lambda_{\textnormal{unimodular}} \bigr\} 
  \subset \bigl\{\Lambda_{\textnormal{isodual}} \bigr\}
  \subset \bigl\{\Lambda_{\textnormal{formally unimodular}}\bigr\}.
  \label{eq:relations_dual-lattices}
\end{IEEEeqnarray*}
\end{remark}

\begin{proposition}
  \label{prop:volume_FU-lattices}
  If $\Lambda$ is formally unimodular, then $\vol{\Lambda}=1$.
\end{proposition}
\ifthenelse{\boolean{short_version}}{}{
\begin{IEEEproof}
  Since by definition $\Theta_{\Lambda}(z)=\Theta_{\Lambda^\star}(z)$,~\eqref{eq:Jacobi-formula} becomes
\begin{IEEEeqnarray}{c}
  \Theta_{\Lambda}(z)
  = \vol{\Lambda^\star}\Bigl(\frac{i}{z}\Bigr)^{\frac{n}{2}}\Theta_{\Lambda}\Bigl(-\frac{1}{z}\Bigr). 
  \label{eq:Jacobi-formula_FU}
\end{IEEEeqnarray}
Also, applying~\eqref{eq:Jacobi-formula} to the dual lattice yields
\begin{IEEEeqnarray}{c}
  \Theta_{\Lambda^\star}(z)=\vol{\Lambda}\Bigl(\frac{i}{z}\Bigr)^{\frac{n}{2}}\Theta_{\Lambda}\Bigl(-\frac{1}{z}\Bigr).
  \label{eq:Jacobi-formula_dualFU}
\end{IEEEeqnarray}
By comparing \eqref{eq:Jacobi-formula_FU} with \eqref{eq:Jacobi-formula_dualFU}, we have $\vol{\Lambda}=\vol{\Lambda^\star}$ because of $\Theta_{\Lambda}(z)=\Theta_{\Lambda^\star}(z)$. It then follows from Remark~\ref{rm:volume} that $\vol{\Lambda}=1$.
\end{IEEEproof}
}
Consequently, unimodular, isodual, and formally unimodular lattices satisfy
\begin{IEEEeqnarray}{c}
  \Theta_{\Lambda}(z)=\Bigl(\frac{i}{z}\Bigr)^{\frac{n}{2}}\Theta_{\Lambda}\Bigl(-\frac{1}{z}\Bigr).
  \label{eq:FSD_Jacobi-formula}
\end{IEEEeqnarray}

Lattices can be constructed from linear codes through the so called~\emph{Construction A}.	 
\begin{definition}[Construction A]
  Let $\code{C}$ be an $[n,k]$ code, then
  \begin{IEEEeqnarray*}{c}
    \ConstrA{\code{C}} \eqdef \tfrac{1}{\sqrt{2}}\left(\phi(\code{C}) + 2\mathbb{Z}^n\right),
  \end{IEEEeqnarray*}
  is a lattice, where $\phi: \Field_2^n \rightarrow \Reals^n$ is the natural embedding.
\end{definition}

About Construction A lattices obtained from codes over $\mathbb{F}_2,$ it is known from~\cite[p.~183]{ConwaySloane99_1} that
\begin{itemize}
\item The volume is $\vol{\ConstrA{\code{C}}} = \frac{2^{\nicefrac{n}{2}}}{|\code{C}|} = 2^{\nicefrac{(n-2k)}{2}}$.
\item $\ConstrA{\code{C}^{\perp}}=\ConstrA{\code{C}}^\star$. 
\end{itemize}

A connection between the weight enumerator $W_C(x,y)$ of a code $\code{C}$ and a lattice $\ConstrA{\code{C}}$ can be established.
\begin{lemma}[{\cite[Th.~3, Ch.~7]{ConwaySloane99_1}}]
  \label{lem:ThetaSeries_WeightDistribution_ConstructionA}
  Consider an $[n,k]$ code $\code{C}$ with $W_{\code{C}}(x,y)$, then the theta series of $\ConstrA{\code{C}}$ is given by
  \begin{IEEEeqnarray*}{c}
    \Theta_{\Lambda_\textnormal{A}(\code{C})}(z) = W_{\code{C}}(\vartheta_3(2z), \vartheta_2(2z)).
  \end{IEEEeqnarray*}
\end{lemma}

\begin{remark}
  \label{rmk:FSD-codes_ConstructionA}
  It follows immediately from Lemma~\ref{lem:ThetaSeries_WeightDistribution_ConstructionA} that if an $[n,\nicefrac{n}{2}]$ code $\code{C}$ is formally self-dual then $\ConstrA{\code{C}}$ is a formally unimodular lattice.
\end{remark}

\section{Secrecy Function of a Lattice}
\label{sec:secrecy-function_lattice}

In the Gaussian WTC, the same coset encoding idea proposed in Wyner's seminal paper~\cite{Wyner75_1} for linear codes can be implemented in a lattice scenario, and here we follow the lattice coding scheme proposed in ~\cite{BelfioreSole10_1,OggierSoleBelfiore16_1}. 

In practice, two lattices $\Lambda_\textnormal{e} \subset \Lambda_\textnormal{b}$ are considered. $\Lambda_\textnormal{b}$ is designed to ensure reliability for a legitimate receiver Bob and required to have a good \emph{Hermite parameter} 
(that measures the highest attainable coding gain of an $n-$dimensional lattice)~\cite{ConwaySloane99_1}. On the other hand, $\Lambda_\textnormal{e}$ is aimed to increase the eavesdropper confusion, so it should be chosen such that $P_{c,\textnormal{e}}$, the eavesdropper's success probability of correctly guessing the transmitted message, is minimized. The performance of the lattice $\Lambda_e$ is measured in terms of the secrecy gain~\cite{BelfioreSole10_1,OggierSoleBelfiore16_1}; to be explained next.
	

Denote by $\sigma_{\textnormal{e}}^2$ the variance of the additive Gaussian noise at the eavesdropper's side. Minimizing $P_{c,\textnormal{e}}$ is equivalent to~\cite{OggierSoleBelfiore16_1} minimizing
\begin{IEEEeqnarray*}{c}
  \sum_{\vect{r}\in\Lambda_\textnormal{e}} e^{-\nicefrac{\norm{\vect{r}}^2}{2 \sigma_e^2}} = \Theta_{\Lambda_\textnormal{e}} \biggl(z\eqdef\frac{i}{2\pi\sigma_\textnormal{e}^2} \biggr),
\end{IEEEeqnarray*}
subject to $\log_2\card{\nicefrac{\Lambda_\textnormal{b}}{\Lambda_\textnormal{e}}}=k$. Note that $\bigIm{\nicefrac{i}{2\pi\sigma_\textnormal{e}^2}} =\Im{z}>0$, thus we consider only 
the positive values of $\tau\eqdef-iz=\nicefrac{1}{2\pi\sigma_e^2}>0$ for $\Theta_{\Lambda_\textnormal{e}}(z)$. Hence, the scheme is aimed at finding a good lattice $\Lambda_{\textnormal{e}}$ such that $\Theta_{\Lambda_\textnormal{e}}(z)$ is minimized, which motivates the following definition of the secrecy function.
\begin{definition}[Secrecy function and secrecy gain~{\cite[Def.~1 and~2]{OggierSoleBelfiore16_1}}]
  Let $\Lambda$ be a lattice with volume $\vol{\Lambda}=\nu^n$. The secrecy function of $\Lambda$ is defined by
  \begin{IEEEeqnarray*}{c}
    \Xi_{\Lambda}(\tau)\eqdef\frac{\Theta_{\nu\Integers^n}(i\tau)}{\Theta_{\Lambda}(i\tau)},
    \label{eq:def_secrecy-function}
  \end{IEEEeqnarray*} 
  for $\tau\eqdef -i z>0$. As maximizing $\Xi_{\Lambda}(\tau)$ is equivalent to minimizing $\Theta_{\Lambda}(z)$, the \emph{(strong) secrecy gain} of a lattice is given by
  $\xi_{\Lambda}\eqdef\sup_{\tau>0}\Xi_{\Lambda}(\tau)$.
\end{definition}

Ideally, the goal is to determine $\xi_{\Lambda}$. However, since the global maximum of a secrecy function is in general not always easy to calculate, a weaker definition is useful. We start by defining the \emph{symmetry point}.
\begin{definition}[Symmetry point]
  A point $\tau_0\in \Reals$ is said to be a \emph{symmetry point} if for all $\tau >0$,
  \begin{IEEEeqnarray}{c}
    \Xi(\tau_0 \cdot \tau) = \Xi\Bigl(\frac{\tau_0}{\tau}\Bigr).
    \label{eq:symmetry-point_tau0}
  \end{IEEEeqnarray}
\end{definition}

\begin{definition}[Weak secrecy gain~{\cite[Def.~3]{OggierSoleBelfiore16_1}}]
  If the secrecy function of a lattice $\Lambda$ has a symmetry point $\tau_0$, then the weak secrecy gain $\chi_\Lambda$ is defined as $\chi_\Lambda=\Xi_\Lambda(\tau_0)$.
\end{definition}

\section{Weak Secrecy Gain of Formally Unimodular Lattices}
\label{sec:weak-secrecy-gain_FU}

This section shows that formally unimodular lattices also hold the same secrecy function properties as unimodular and isodual lattices~\cite{OggierSoleBelfiore16_1}.
\begin{lemma}
  \label{lem:secrecy_lattice_dual}
  Consider a lattice $\Lambda$ and its dual $\Lambda^\star$. Then,
  \begin{IEEEeqnarray}{c}
    \Xi_\Lambda(\tau) = \Xi_{\Lambda^\star}\Bigl(\frac{1}{\tau}\Bigr).
    \label{eq:secrecy-functions_dual-lattices}
  \end{IEEEeqnarray}
\end{lemma}
\ifthenelse{\boolean{short_version}}{}{
\begin{IEEEproof}
  Recall the scaling properties of the theta series: for any $c\in\Reals$, we have $\Theta_{c\Lambda}(z) = \Theta_{\Lambda}(c^2 z)$. Therefore, 
\begin{IEEEeqnarray*}{rCl}
\Xi_\Lambda(\tau) & = & \frac{\Theta_{\nu \mathbb{Z}^n}(i\tau)}{\Theta_{\Lambda}(i\tau)} = \frac{\Theta_{\mathbb{Z}^n}\bigl(\nu^2 \cdot i\tau\bigr)}{\Theta_{\Lambda}(i\tau)}\\
& \stackrel{\text{\eqref{eq:Jacobi-formula}}}{=} & \frac{\vol{\Integers^n}(\nu^2\tau)^{-\nicefrac{n}{2}}\cdot\Theta_{\mathbb{Z}^n}\bigl(\frac{i}{\nu^2\tau}\bigr)}{\vol{\Lambda^\star}\tau^{-n/2}\cdot\Theta_{\Lambda^\star}\bigl(\frac{i}{\tau}\bigr)}
\\
& \stackrel{(a)}{=} & \frac{\Theta_{\mathbb{Z}^n}\bigl(\frac{i}{\nu^2\tau}\bigr)}{\Theta_{\Lambda^\star}\bigl(\frac{i}{\tau}\bigr)} =  \dfrac{\Theta_{\inv{\nu} \mathbb{Z}^n}\left(\frac{i}{\tau} \right)}{\Theta_{\Lambda^\star}\left(\frac{i}{\tau} \right)} \stackrel{(b)}{=} \Xi_{\Lambda^\star}\Bigl(\frac{1}{\tau}\Bigr),
\end{IEEEeqnarray*}
where $(a)$ and $(b)$ hold since $\vol{\Lambda^\star}=\inv{\vol{\Lambda}}=\nu^{-n}$. 
\end{IEEEproof}
}

A necessary and sufficient condition for a lattice $\Lambda$ 
to achieve the weak secrecy gain at $\tau=1$ is given as follows.

\begin{theorem}
  \label{thm:weak_secrecy}
  Consider a lattice $\Lambda$ with $\vol{\Lambda}=1$ and its dual $\Lambda^\star$.
  Then, $\Lambda$ achieves the weak secrecy gain at $\tau=1$, if and only if $\Lambda$ is formally unimodular.
\end{theorem}
\begin{IEEEproof}
  By definition, we have
  \begin{IEEEeqnarray}{c}
    \Xi_\Lambda(\tau) = \Xi_\Lambda\Bigl(\frac{1}{\tau}\Bigr).
    \label{eq:symmetry-point_at-tau1}
  \end{IEEEeqnarray}
  Using Lemma~\ref{lem:secrecy_lattice_dual}, it follows 
  from~\eqref{eq:symmetry-point_at-tau1} and~\eqref{eq:secrecy-functions_dual-lattices} that 
\begin{IEEEeqnarray*}{c}
  \Xi_{\Lambda}\Bigl(\frac{1}{\tau}\Bigr)=\Xi_\Lambda(\tau)
  =\Xi_{\Lambda^\star}\Bigl(\frac{1}{\tau}\Bigr).
\end{IEEEeqnarray*}
By Def.~\ref{eq:def_secrecy-function}, this implies that $\Theta_{\Lambda}(z)=\Theta_{\Lambda^\star}(z)$ 
for $\vol{\Lambda}=1$. 
Conversely, from Def.~\ref{def:uni-iso-fum}, we
see that~\eqref{eq:secrecy-functions_dual-lattices} implies~\eqref{eq:symmetry-point_at-tau1}.
\end{IEEEproof}
Note that Theorem~\ref{thm:weak_secrecy} holds for isodual lattices as well, which yields to \cite[Prop.~1]{OggierSoleBelfiore16_1}.


\begin{corollary}
  \label{cor:weak_secrecy_not-at-tau1}
  Consider a lattice $\Lambda$ with $\vol{\Lambda}=\nu^n$ and its dual $\Lambda^\star$. Then, $\Lambda$ achieves the weak secrecy gain at $\tau=\nu^{-2}$, 
  if and only if $\inv{\nu}\Lambda$ is a formally unimodular lattice.
\end{corollary}
\ifthenelse{\boolean{short_version}}{}{
\begin{IEEEproof}
  Consider a lattice $\tilde{\Lambda}= \inv{\nu}\Lambda$. 
  Then, observe that
  \begin{IEEEeqnarray*}{rCl}
    \Xi_{\Lambda}(\nu^{-2}  \cdot \tau) & = & \frac{\Theta_{\nu \mathbb{Z}^n}(\nu^{-2} \cdot i \tau)}{\Theta_{\Lambda}(\nu^{-2} \cdot i \tau)} = \frac{\Theta_{\mathbb{Z}^n}(i\tau)}{\Theta_{\inv{\nu} \Lambda}(i\tau)}=\frac{\Theta_{\mathbb{Z}^n}(i \tau)}{\Theta_{\tilde{\Lambda}}(i \tau)} 
  \\
  & = &\Xi_{\tilde{\Lambda}}(\tau), \textnormal{ and}
  \\
  \Xi_\Lambda\Bigl(\frac{\nu^{-2}}{\tau}\Bigr)& = &\Xi_{\tilde{\Lambda}}\Bigl(\frac{1}{\tau}\Bigr).
\end{IEEEeqnarray*}
Direct application of Theorem~\ref{thm:weak_secrecy} completes the proof.
\end{IEEEproof}
}

Equation~\eqref{eq:symmetry-point_tau0} with $\tau_0=\nu^{-2}$ holds for a lattice equivalent to its dual. See~\cite[Prop.~2]{OggierSoleBelfiore16_1}.

\section{Secrecy Gain of Formally Unimodular Lattices}
\label{sec:secrecy-gain_FU-lattices}
Our goal in this section is to investigate the following conjecture.
\begin{conjecture}
  \label{conj:secrecy-gain_FU-lattices}
  The secrecy function of a formally unimodular lattice $\Lambda$ achieves its maximum at $\tau=1$, i.e., $\xi_{\Lambda}=\Xi_{\Lambda}(1)$.
\end{conjecture}

Although we cannot completely prove Conjecture~\ref{conj:secrecy-gain_FU-lattices}, we proceed to study the 
secrecy gain for formally unimodular lattices obtained from formally self-dual codes via Construction A (see Remark~\ref{rmk:FSD-codes_ConstructionA}). Note that for linear codes, it is known that formally self-dual codes that are not self-dual can outperform self-dual codes in some cases, as they comprise a wider class and hence may allow a better minimum Hamming distance or an overall more favorable weight enumerator. This leads us to look for 
improved results on the secrecy gain compared to unimodular lattices~\cite{Ernvall-Hytonen12_1,LinOggier13_1,Pinchak13_1}.
\begin{lemma}
  \label{lem:ThetaSeries_ConstructionA_FSDcodes}
  Consider a Construction A lattice $\Lambda_{\textnormal{A}}(\code{C})$ obtained from a formally self-dual code $\code{C}$. Then, its theta series is equal to
  \begin{IEEEeqnarray*}{c}
    \Theta_{\Lambda_{\textnormal{A}}(\code{C})}=\frac{W_{\code{C}}\left(\sqrt{\vartheta^2_3(z)+\vartheta^2_4(z)},\sqrt{\vartheta^2_3(z)-\vartheta^2_4(z)}\right)}{{2^{\frac{n}{2}}}}.
  \end{IEEEeqnarray*}
\end{lemma}
\begin{IEEEproof}
  Using Lemma~\ref{lem:ThetaSeries_WeightDistribution_ConstructionA} and the useful identities given in~\cite[eq.~(26), Ch.~4]{ConwaySloane99_1}, the theta series $\Theta_{\Lambda_{\textnormal{A}}(\code{C})}$ becomes 
  \begin{IEEEeqnarray*}{rCl}
    \IEEEeqnarraymulticol{3}{l}{%
      \Theta_{\Lambda_{\textnormal{A}}(\code{C})}(z)}\nonumber\\*\quad%
    & = &W_{\code{C}}(\vartheta_3(2z),\vartheta_2(2z))
    \\
    & \stackrel{(a)}{=} &W_{\code{C}}\left(\frac{\vartheta_3(2z)+\vartheta_2(2z)}{\sqrt{2}},\frac{\vartheta_3(2z)-\vartheta_2(2z)}{\sqrt{2}}\right)
    \\
    & = &W_{\code{C}}\Biggl(\frac{\sqrt{\vartheta^2_3(z)+\vartheta^2_4(z)}+\sqrt{\vartheta^2_3(z)-\vartheta^2_4(z)}}{\sqrt{2}\sqrt{2}},\nonumber\\
    &&\qquad\quad\>\frac{\sqrt{\vartheta^2_3(z)+\vartheta^2_4(z)}-\sqrt{\vartheta^2_3(z)-\vartheta^2_4(z)}}{\sqrt{2}\sqrt{2}}\Biggr)
    \\
    & = &\frac{1}{2^{\frac{n}{2}}}W_{\code{C}}\Biggl(\frac{\sqrt{\vartheta^2_3(z)+\vartheta^2_4(z)}+\sqrt{\vartheta^2_3(z)-\vartheta^2_4(z)}}{\sqrt{2}},\nonumber\\
    &&\qquad\qquad\>\frac{\sqrt{\vartheta^2_3(z)+\vartheta^2_4(z)}-\sqrt{\vartheta^2_3(z)-\vartheta^2_4(z)}}{\sqrt{2}}\Biggr)
    \\
    & \stackrel{(b)}{=} &\frac{1}{2^{\frac{n}{2}}}W_{\code{C}}\left(\sqrt{\vartheta^2_3(z)+\vartheta^2_4(z)},\sqrt{\vartheta^2_3(z)-\vartheta^2_4(z)}\right).
  \end{IEEEeqnarray*}
  where $(a)$ and $(b)$ follow from~\eqref{eq:FSD_MacWilliams-identity}.
\end{IEEEproof}
\begin{lemma}
  \label{lem:ratio_theta4-theta3}
  Let $s(\tau)\eqdef\nicefrac{\vartheta_4(i\tau)}{\vartheta_3(i\tau)}$. Then,
  $s(\tau)$ is an increasing function for $\tau>0$, and $0< s(\tau) < 1$.
\end{lemma}
\ifthenelse{\boolean{short_version}}{}{
\begin{IEEEproof}
  The detailed proof is given in Appendix~\ref{sec:proof_ratio_theta4-theta3}.
\end{IEEEproof}
}
\begin{remark}
  \label{rem:bijective_t-tau}
  Let $t(\tau)\eqdef s(\tau)^2$. Then, $0<t(\tau) < 1$ and $t(\tau)$ is also an increasing function for $\tau>0$. Hence, according to Lemma~\ref{lem:ratio_theta4-theta3}, given any $t\in(0,1)$, there always exists a unique $\tau>0$ such that $t(\tau)=\nicefrac{\vartheta^2_4(i\tau)}{\vartheta^2_3(i\tau)}$. Moreover, we have $t(1)=\nicefrac{1}{\sqrt{2}}$ by using the identity of $\vartheta_3(i)=2^{\nicefrac{1}{4}}\vartheta_4(i)$ from~\cite{Weisstein_3}. 
\end{remark}
From Remark~\ref{rem:bijective_t-tau} and Lemma~\ref{lem:ThetaSeries_ConstructionA_FSDcodes}, now we are able to give a new universal approach to derive the strong secrecy gain of a Construction A lattice obtained from formally self-dual codes.
\begin{theorem}
  \label{thm:inv_secrecy-function_WeightEnumerator}
  Let $\code{C}$ be a formally self-dual code. Then
  \begin{IEEEeqnarray*}{c}
    \inv{\left[\Xi_{\Lambda_\textnormal{A}(\code{C})}(\tau)\right]}=\frac{W_{\code{C}}\bigl(\sqrt{1+t(\tau)},\sqrt{1-t(\tau)}\bigr)}{2^\frac{n}{2}},\label{eq:Xi-ft_ConstructionA_FSDcodes}\IEEEeqnarraynumspace
  \end{IEEEeqnarray*}
  where $0<t(\tau)=\nicefrac{\vartheta_4^2(i\tau)}{\vartheta^2_3(i\tau)} < 1$. Moreover, define $f_{\code{C}}(t)\eqdef W_\code{C}(\sqrt{1+t},\sqrt{1-t})$ for $0< t < 1$. Then, maximizing the secrecy function $\Xi_{\ConstrA{\code{C}}}(\tau)$ is equivalent to determining the minimum of $f_{\code{C}}(t)$ on $t\in(0,1)$.
\end{theorem}
\begin{example}
  \label{ex:ex_n6k3d3}
  Consider a $[6,3,3]$ odd formally self-dual code $\code{C}$ with $W_{\code{C}}(x,y) = x^6+4x^3y^3+3x^2y^4$~\cite{BetsumiyaHarada01_1}. Thus $f_{\code{C}}(t) = W_{\code{C}}(\sqrt{1+t},\sqrt{1-t})=4[1+t^3+(1-t^2)^{\nicefrac{3}{2}}]$ and $f'_{\code{C}}(t)=12t(t-\sqrt{1-t^2})$. Observe that for $0<t<\nicefrac{1}{\sqrt{2}}$, we have $\sqrt{1-t^2}>\nicefrac{1}{\sqrt{2}}$. Then, $t-\sqrt{1-t^2}<\nicefrac{1}{\sqrt{2}}-\nicefrac{1}{\sqrt{2}}=0$. This indicates that the derivative $f'_{\code{C}}(t)<0$ on $t\in(0,\nicefrac{1}{\sqrt{2}})$.
Similarly, one can also show that $f_{\code{C}}'(t)>0$ on $t\in(\nicefrac{1}{\sqrt{2}},1)$, and $t=\nicefrac{1}{\sqrt{2}}$ is the minimum of $f_{\code{C}}(t)$. Hence, Remark~\ref{rem:bijective_t-tau} and Theorem~\ref{thm:inv_secrecy-function_WeightEnumerator} indicate that the maximum of $\Xi_{\ConstrA{\code{C}}}(\tau)$ is achieved at $\tau=1$. Also, one can get $\xi_{\ConstrA{\code{C}}}\approx 1.172$.\hfill\exampleend

\end{example}   

\ifthenelse{\boolean{short_version}}{Using Gleason's Theorem~\cite[Th.~9.2.1]{HuffmanPless03_1}, an expression of $f_{\code{C}}(t)$ can be shown if $\code{C}$ is an even formally self-dual code.}{The following lemma shows a general expression of $f_{\code{C}}(t)$ if $\code{C}$ is an even formally self-dual code.}
\begin{lemma}
  \label{lem:f_C_even-FSDcodes}
  If $\code{C}$ is an $[n,\nicefrac{n}{2}]$ even formally self-dual codes, then we have
  \begin{IEEEeqnarray}{c}
    f_{\code{C}}(t)=2^{\frac{n}{2}}\sum_{r=0}^{\lfloor\frac{n}{8}\rfloor}a_r (t^4-t^2+1)^r,
    \label{eq:ThetaSeries_evenFSD-codes}
  \end{IEEEeqnarray}
  where $a_r\in\Rationals$ and $\sum_{r=0}^{\lfloor\frac{n}{8}\rfloor}a_r=1$.
\end{lemma}
\ifthenelse{\boolean{short_version}}{}{
\begin{IEEEproof}
  Consider $g_1(x,y)=x^2+y^2$ and $g_2(x,y)=x^8+14x^4y^4+y^8$. Then, by performing some simple calculations, we obtain
  \begin{IEEEeqnarray*}{rCl}
    g_1(\sqrt{1+t},\sqrt{1-t})& = &2,
    \\
    g_2(\sqrt{1+t},\sqrt{1-t})& = &16(t^4-t^2+1).
  \end{IEEEeqnarray*}
  Therefore,~\eqref{eq:ThetaSeries_evenFSD-codes} follows from Gleason's Theorem~\cite[Th.~9.2.1]{HuffmanPless03_1}.
\end{IEEEproof}
}

Next, we provide a sufficient condition for a Construction A formally unimodular lattice obtained from even formally self-dual codes to achieve the strong secrecy gain at $\tau=1$, or, equivalently, $t=\nicefrac{1}{\sqrt{2}}$.
\begin{theorem}
  \label{thm:strong-secrecy-gain_unimodular-lattices}
  Consider $n \geq 8$ and an $[n,\nicefrac{n}{2}]$ even formally self-dual code $\code{C}$. If the coefficients $a_r$ of $f_{\code{C}}(t)$ expressed in terms of \eqref{eq:ThetaSeries_evenFSD-codes} satisfy
  \begin{IEEEeqnarray}{c}
    \sum_{r=1}^{\lfloor\frac{n}{8}\rfloor} r a_r\Bigl(\frac{3}{4}\Bigr)^{r-1} > 0,
    \label{eq:condition_ai_evenFSD}
  \end{IEEEeqnarray}
  then the secrecy gain of $\ConstrA{\code{C}}$ is achieved at $\tau=1$.
\end{theorem}
\begin{IEEEproof}
  It is enough to show that the function $f_{\code{C}}(t)$ as in \eqref{eq:ThetaSeries_evenFSD-codes} defined for $0 < t < 1$ achieves its 
  minimum at $t=\nicefrac{1}{\sqrt{2}}$. 
  
  Since $h(t)\eqdef t^4-t^2+1 = (t^2-\nicefrac{1}{4})^2+\nicefrac{3}{4}\geq\nicefrac{3}{4}$ on $t\in(0,1)$, the derivative of $f_{\code{C}}(t)$ satisfies
  \begin{IEEEeqnarray*}{c}
    \frac{\dd f_{\code{C}}(t)}{\dd t} = 2^{\frac{n}{2}} h'(t) \sum_{r=1}^{\lfloor\frac{n}{8}\rfloor} r a_r h(t)^{r-1} \geq 2^{\frac{n}{2}} h'(t) \sum_{r=1}^{\lfloor\frac{n}{8}\rfloor} r a_r \Bigl( \frac{3}{4}\Bigr)^{r-1}
  \end{IEEEeqnarray*}  
  and $h'(t)=4t^3-2t=2t(2t^2-1)$. As the hypothesis holds, the behavior of the derivative is dominated by $h'(t)$. Since
  \begin{IEEEeqnarray*}{c}
    h'(t)
    \begin{cases}
      <0 & \textnormal{if }0< t<\frac{1}{\sqrt{2}},
      \\
      =0 & \textnormal{if }t=\frac{1}{\sqrt{2}},
      \\
      >0 & \textnormal{if }\frac{1}{\sqrt{2}}< t < 1,
    \end{cases},
  \end{IEEEeqnarray*}
  it implies that $f_{\code{C}}(t)$ is decreasing in $t\in(0,\nicefrac{1}{\sqrt{2}})$ and increasing in $t\in(\nicefrac{1}{\sqrt{2}},1)$. This completes the proof.
\end{IEEEproof}

\begin{table*}[t]
  \centering
  \caption{Comparison of (strong) secrecy gains for several values of even dimensions $n$. Codes without references are obtained by tailbiting the rate $\nicefrac{1}{2}$ convolution codes.
  }
  \label{tab:table_secrecy-gains_FU-lattices}
  \vskip -2.0ex
  \Scale[1.0]{\begin{IEEEeqnarraybox}[
    \IEEEeqnarraystrutmode
    \IEEEeqnarraystrutsizeadd{3.5pt}{3.0pt}]{V/c/V/c/V/c/V/c/V/c/V/c/V/c/V}
    \IEEEeqnarrayrulerow\\
    & n
    && \code{C}_\textnormal{sd}^{(d)}
    && \xi_{\Lambda_\textnormal{A}(\code{C}_\textnormal{sd})}
    && \code{C}_\textnormal{efsd}^{(d)}
    && \xi_{\Lambda_\textnormal{A}(\code{C}_\textnormal{efsd})}
    && \code{C}_\textnormal{ofsd}^{(d)}
    && \xi_{\Lambda_\textnormal{A}(\code{C}_\textnormal{ofsd})}
    &\\    
    \hline\hline
    &6 && -  && - && \code{C}_\textnormal{efsd}^{(2)}~ \textnormal{\cite{HuffmanPless03_1}} && 1 && \code{C}_\textnormal{ofsd}^{(3)}~\textnormal{\cite{BetsumiyaHarada01_1}} && ~\mathbf{1.172} &\\
    \IEEEeqnarrayrulerow\\
    &8 && \code{C}_\textnormal{sd}^{(4)}~\textnormal{\cite{HuffmanPless03_1}}  && \mathbf{1.333} && - && - && \code{C}_\textnormal{ofsd}^{(3)}~\textnormal{\cite{BetsumiyaHarada01_1}} && ~1.282 &\\
    \IEEEeqnarrayrulerow\\
    &10 && - && - && \code{C}_\textnormal{efsd}^{(4)}~\textnormal{\cite{KennedyPless94_1}}  && 1.455 && \code{C}_\textnormal{ofsd}^4~\textnormal{\cite{BetsumiyaHarada01_1}} && \mathbf{1.478} &\\  
    \IEEEeqnarrayrulerow\\
    &12 && \code{C}_\textnormal{sd}^{(4)}~\textnormal{\cite{LinOggier13_1}} && 1.6 && \code{C}_\textnormal{efsd}^{(4)}~\textnormal{\cite{BetsumiyaGulliverHarada99_1}}  && 1.6 && \code{C}_\textnormal{ofsd}^{(4)}~\textnormal{\cite{BetsumiyaHarada01_1}}  && \mathbf{1.657} & \\
    \IEEEeqnarrayrulerow\\
    &14 && \code{C}_\textnormal{sd}^{(4)}~\textnormal{\cite{LinOggier13_1}} && 1.778 && \code{C}_\textnormal{efsd}^{(4)}~\textnormal{\cite{BetsumiyaGulliverHarada99_1}} && 1.825 && \code{C}_\textnormal{ofsd}^{(4)}~\textnormal{\cite{BetsumiyaHarada01_1}}  && \mathbf{1.875} & \\
    \IEEEeqnarrayrulerow\\
    &16 && \code{C}_\textnormal{sd}^{(4)}~\textnormal{\cite{LinOggier13_1}} && 2 && \code{C}_\textnormal{efsd}^{(4)}~\textnormal{\cite{BetsumiyaHarada01_2}}  && 2.133 && \code{C}_\textnormal{ofsd}^{(5)}~\textnormal{\cite{BetsumiyaHarada01_1}}  && \mathbf{2.141} & \\
    \IEEEeqnarrayrulerow\\
    &18 && \code{C}_\textnormal{sd}^{(4)}~\textnormal{\cite{LinOggier13_1}} && 2.286 && \code{C}_\textnormal{efsd}^{(6)}~\textnormal{\cite{SloaneHeninger06_1}}   && \mathbf{2.485} && \code{C}_\textnormal{ofsd}^{(5)}  && 2.427 & \\
    \IEEEeqnarrayrulerow\\
    &20 && \code{C}_\textnormal{sd}^{(4)}~\textnormal{\cite{LinOggier13_1}} && 2.523 && \code{C}_\textnormal{efsd}^{(6)}~ \textnormal{\cite{FieldsGaboritHuffmanPless01_1}}  && 2.813 && \code{C}_\textnormal{ofsd}^{(6)}~\textnormal{\cite{BetsumiyaGulliverHarada99_1}}   && \mathbf{2.868} & \\
    \IEEEeqnarrayrulerow\\
    &22 && \code{C}_\textnormal{sd}^{(6)}~\textnormal{\cite{LinOggier13_1}} && 3.2 &&  \code{C}_\textnormal{efsd}^{(6)}  && 3.2  && \code{C}_\textnormal{ofsd}^{(7)}~\textnormal{\cite{BetsumiyaHarada01_1}}  && \mathbf{3.335} & \\
    \IEEEeqnarrayrulerow\\
     &30 && \code{C}_\textnormal{sd}^{(6)}~\textnormal{\cite{ConwaySloane90_1}} && 5.697 && \code{C}_\textnormal{efsd}^{(8)}~\textnormal{\cite{BouyuklievaBouyukliev10_1}}   && \mathbf{5.843} && \code{C}_\textnormal{ofsd}^{(7)}  && 5.785 & \\
    \IEEEeqnarrayrulerow\\
    &32 && \code{C}_\textnormal{sd}^{(8)}~\textnormal{\cite{ConwaySloane90_1}} && 6.737 && 
    {\code{C}}_\textnormal{efsd}^{(8)}
    && \mathbf{6.748} && \code{C}_\textnormal{ofsd}^{(7)}  && 6.628 & \\
    \IEEEeqnarrayrulerow\\
    &40 && \code{C}_\textnormal{sd}^{(8)}~\textnormal{\cite{ConwaySloane90_1}} && 12.191 && 
    {\code{C}}_\textnormal{efsd}^{(8)}
    && 12.134 && \code{C}_\textnormal{ofsd}^{(9)}  && \mathbf{12.364} & \\
    \IEEEeqnarrayrulerow \\
    &70 && \code{C}_\textnormal{sd}^{(12)}~\textnormal{\cite{Harada97_1}} && 127.712 && 
    {\code{C}}_\textnormal{efsd}^{(12)}
    && 128.073 && \code{C}_\textnormal{ofsd}^{(13)}  && \mathbf{128.368} & \\
    \IEEEeqnarrayrulerow
  \end{IEEEeqnarraybox}}
\end{table*}

\begin{example}
  \label{ex:n18k9d6}
  Consider an $[18,9,6]$ even formally self-dual code $\code{C}$ with
  \begin{IEEEeqnarray*}{rCl}
    W_{\code{C}}(x,y) & = &x^{18}+102 x^{12} y^6+153 x^{10} y^8
    \nonumber\\
    && \>+153 x^8 y^{10}+102 x^6 y^{12}+y^{18}. 
  \end{IEEEeqnarray*}
  By solving $f_{\code{C}}(t)=W_{\code{C}}(\sqrt{1+t},\sqrt{1-t})$ with~\eqref{eq:ThetaSeries_evenFSD-codes} (see the details of derivation provided in~\ifthenelse{\boolean{short_version}}{~\cite[App.~B]{BollaufLinYtrehus21_1sub}}{Appendix~\ref{sec:coefficients_gleason}}), we find that $a_0=-\nicefrac{29}{16}, a_1=\nicefrac{27}{8}$ and $a_2=-\nicefrac{9}{16}$. The condition~\eqref{eq:condition_ai_evenFSD} in Theorem~\ref{thm:strong-secrecy-gain_unimodular-lattices} for those coefficients is satisfied since $\nicefrac{27}{8}-\nicefrac{27}{32}=\nicefrac{81}{32}>0$. Thus, the secrecy gain conjecture is true for the formally unimodular lattice $\ConstrA{\code{C}}$.\hfill\exampleend
\end{example}

\section{Numerical Results}
\label{sec:numerical-results}

Even though the result of Theorem~\ref{thm:strong-secrecy-gain_unimodular-lattices} is restricted to formally unimodular lattices obtained from even formally self-dual codes, we have numerical evidence showing that Conjecture~\ref{conj:secrecy-gain_FU-lattices} also holds for formally unimodular lattices obtained from odd formally self-dual codes. The secrecy gains of some formally unimodular Construction A lattices obtained from (even and odd) formally self-dual codes are summarized in Table~\ref{tab:table_secrecy-gains_FU-lattices}. Note that all codes have the parameters $[n, \nicefrac{n}{2}]$ and the superscript ``$(d)$'' refers to the minimum Hamming distance $d$ of the code. Their exact weight enumerators can be found in~\ifthenelse{\boolean{short_version}}{~\cite[App.~D]{BollaufLinYtrehus21_1sub}}{Appendix~\ref{sec:weight-enumerators-codes_TableI}}. The highlighted values represent the best values found in the respective dimensions, when comparing self-dual (sd), even and odd formally self-dual (efsd and ofsd) codes.

\begin{remark}
  \label{rem:remark_TableI}
  We remark the following about Table~\ref{tab:table_secrecy-gains_FU-lattices}:
  \begin{itemize}
    \item ``[$\cdot$]'' indicates the reference number.
    \item We use the sufficient condition~\eqref{eq:condition_ai_evenFSD} in Theorem~\ref{thm:strong-secrecy-gain_unimodular-lattices} for the even codes and the numerical derivative analysis with Wolfram Mathematica~\cite{Mathematica} for the odd codes to confirm the strong secrecy gain in Table~\ref{tab:table_secrecy-gains_FU-lattices}.
    \item For most dimensions $n>8$, the secrecy gain of formally unimodular lattices that are not unimodular exceeds the performance of unimodular lattices (obtained from self-dual codes), presented in~\cite[Tables~I and II]{LinOggier13_1}.  In some cases (\emph{e.g.} [12,6], [22,11]) we were unable to find good efsd codes with different secrecy gains form the sd codes.
    \item Observe that for codes of length $40$, the self-dual code tabulated is a Type I (weights divisible by two), as it presents a higher secrecy gain ($\xi_{\Lambda_\textnormal{A}(\code{C}_\textnormal{sd})} \approx 12.191$) compared to the Type II (weights divisible by four) ($\xi_{\Lambda_\textnormal{A}(\code{C}_\textnormal{sd})} \approx 11.977$). The same happens with codes of length $32$ and this confirms the advantage of this approach as to the results in \cite{Pinchak13_1}.
    \item Formally self-dual (isodual) codes without references in Table~\ref{tab:table_secrecy-gains_FU-lattices} are constructed by tailbiting the rate $\nicefrac{1}{2}$ convolutional codes. Details can be found in~\ifthenelse{\boolean{short_version}}{\cite[App.~C]{BollaufLinYtrehus21_1sub}}{Appendix~\ref{sec:isodual_coco}}.
\end{itemize}
\end{remark}

\section{Conclusion and Future Work}
\label{sec:conclusion}

This paper introduced the \emph{formally unimodular lattices}, a new class consisting of lattices having the same theta series as their dual. We showed some properties of formally unimodular lattices and their secrecy function behavior in the Gaussian WTC. 
Furthermore, we investigated Construction A lattices obtained from formally self-dual codes and gave a universal approach to determine their secrecy gain. We found formally unimodular lattices of better secrecy gain than the best known unimodular lattices from the literature.

The technique we used to construct the theta series of a formally unimodular lattice is based on Construction A from a formally self-dual code. Hence, only results of formally unimodular lattices with even dimensions are discussed. However, 
it is possible to obtain the closed-form expression of the theta series of a formally unimodular lattice with odd dimension, e.g., generalizing Hecke's theorem~\cite[Th.~7, Ch.~7]{ConwaySloane99_1}. This direction of study is of great interest for future research. \ifthenelse{\boolean{short_version}}{}{We also observe that the secrecy gain is generally improved with higher minimum Hamming distance and lower kissing number, and it appears to increase exponentially with the dimension. The precise relation with these parameters will be investigated in a future work.}


\vspace{2ex}

\ifthenelse{\boolean{short_version}}{}{%
\appendices


\section{Proof of Lemma~\ref{lem:ratio_theta4-theta3}}
\label{sec:proof_ratio_theta4-theta3}

By definition, the fact that $0<s(\tau)< 1$ is trivial. Let's directly compute the derivative of $s(\tau)$ and we get
\begin{IEEEeqnarray*}{rCl}
  \frac{\dd t(\tau)}{\dd \tau}& = &\frac{\vartheta_4'(\tau)\vartheta_3(\tau)-\vartheta_4(\tau)\vartheta_3'(\tau)}{\theta_3(\tau)^2}
  \\
  & = &\frac{1}{\theta_3(\tau)^2}\Biggl[\biggl(2\pi\sum_{m=1}^{\infty}(-1)^{m}(-m^2) e^{-\pi\tau(m^2)}\biggr)
  \nonumber\\
  &&\quad\qquad\>\biggl(1+2\sum_{m=1}^{\infty}e^{-\pi\tau(m^2)}\biggr)\nonumber\\
  &&\quad\qquad\>-\biggl(1+2\sum_{m=1}^{\infty}(-1)^{m}e^{-\pi\tau(m^2)}\biggr)\nonumber\\
  &&\quad\qquad\>\biggl(2\pi\sum_{m=1}^{\infty}(-m^2)e^{-\pi\tau(m^2)}\biggr)\Biggr]
  \\
  & = &\frac{1}{\theta_3(\tau)^2}\left[2\pi\sum_{m=1}^{\infty}(-1)^{m+1}m^2 e^{-\pi\tau(m^2)}\right.
  \nonumber\\
  && +2^2\pi\Biggl(\sum_{m=1}^{\infty}(-1)^{m+1}m^2 e^{-\pi\tau(m^2)}\Biggr)\Biggl(\sum_{m=1}^{\infty}e^{-\pi\tau(m^2)}\Biggr)
  \nonumber\\
  &&\>-(-2\pi)\sum_{m=1}^{\infty}m^2e^{-\pi\tau(m^2)}
  \nonumber\\
  && \>-2^2\pi\Biggl(\sum_{m=1}^{\infty}(-1)^{m+1}e^{-\pi\tau(m^2)}\Biggr)\Biggl(\sum_{m=1}^{\infty}m^2 e^{-\pi\tau(m^2)}\Biggr)
  \\
  & = &\frac{1}{\theta_3(\tau)^2}\left[4\pi\sum_{\substack{m=1\\m\colon\textnormal{odd}}}^{\infty}m^2 e^{-\pi\tau(m^2)}\right]>0.
\end{IEEEeqnarray*}
This shows that $s(\tau)$ is increasing on $\tau>0$.

\section{Determining the Coefficients in~\eqref{eq:ThetaSeries_evenFSD-codes} from the Weight Enumerator}
\label{sec:coefficients_gleason}


Let $\code{C}$ be an $[n,\nicefrac{n}{2}]$ even formally self-dual code. Gleason's Theorem~\cite[Th.~9.2.1]{HuffmanPless03_1} states that 
\begin{IEEEeqnarray}{c}
  \label{eq:gleason}
  W_{\code{C}}(x,y)=\sum_{r=0}^{\lfloor\nicefrac{n}{8}\rfloor} a_r g_1(x,y)^{\tfrac{n}{2}-4r}g_2(x,y)^{r},
\end{IEEEeqnarray}
where $g_1(x,y)=x^2+y^2$,  $g_2(x,y)=x^8+14x^4y^4+y^8$, $a_r \in \mathbb{Q}$, and $\sum_{r=0}^{\lfloor \tfrac{n}{8} \rfloor} a_r=1.$

Consider the weight enumerator expressed by 
\begin{IEEEeqnarray}{c}\label{eq:weight_enumerator}
  W_\code{C}(x,y)=\sum_{w=0}^n A_w x^{n-w}y^w.
\end{IEEEeqnarray}
We aim to determine the coefficients $a_r$ in~\eqref{eq:gleason} in terms of $A_w$, $w\in[0:n]$, if the coefficients $A_w$ are known. 

Let's first start to expand $g_1(x,y)^{\frac{n}{2}-4r}$ and $g_2(x,y)^r$. Observe that
\begin{IEEEeqnarray*}{rCl}
g_1(x,y)^{\frac{n}{2}-4r} & = & (x^2+y^2)^{\frac{n}{2}-4r} \nonumber \\
& = & \sum_{j=0}^{\frac{n}{2}-4r} \binom{\nicefrac{n}{2}-4r}{j} (x^2)^{(\frac{n}{2}-4r-j)} (y^2)^{j},
\IEEEeqnarraynumspace
\end{IEEEeqnarray*}
and
\begin{IEEEeqnarray*}{rCl}
  \IEEEeqnarraymulticol{3}{l}{%
    g_2(x,y)^{r}}\nonumber\\*%
  & = &(x^8+14x^4y^4+y^8)^r=[(x^4+7y^4)^2-48y^8]^r
  \nonumber \\
  & = & \sum_{h=0}^{r}\binom{r}{h}\biggl[\sum_{\ell=0}^{2r-2h}\binom{2r-2h}{\ell}(x^4)^{2r-2h-\ell}(7y^4)^{\ell}\biggr](-48y^8)^h.
  \nonumber\IEEEeqnarraynumspace
\end{IEEEeqnarray*}

Given $w\in[0:n]$, by collecting the terms of $y^{2j+8h+4\ell}$ for $2j+8h+4\ell=w$, we get
\begin{IEEEeqnarray}{rCl}
  \IEEEeqnarraymulticol{3}{l}{%
    g_1(x,y)^{\frac{n}{2}-4r} g_2(x,y)^r
  }\nonumber\\*\quad%
  & = &\sum_{\substack{2j+8h+4\ell =w\\ j,h,\ell \in\Integers_{\geq 0}}} 7^\ell(-48)^h\binom{\nicefrac{n}{2}-4r}{j}\nonumber \\
  \nonumber\\
  && \>\times\binom{r}{h}\binom{2r-2h}{\ell}x^{n-2j-8h-4\ell} y^{2j+8h+4\ell},
  \label{eq:expand_multiplication}
\end{IEEEeqnarray}
where 
we define $\binom{p}{q}=0,$ if $p<q$.

By comparing the coefficients of~\eqref{eq:weight_enumerator} and~\eqref{eq:gleason}, we get 
\begin{IEEEeqnarray}{rCl}
  A_w & = & \sum_{r=0}^{\lfloor\nicefrac{n}{8} \rfloor} a_r  \sum_{\substack{2j+8h+4\ell =w \\ j,k,\ell \in \Integers_{\geq 0}}} 7^\ell(-48)^h \binom{\nicefrac{n}{2}-4r}{j}\binom{r}{h}\nonumber \\
  &&\>\times\binom{2r-2h}{\ell} x^{n-2j-8h-4\ell} y^{2j+8h+4\ell}.
  \label{eq:Aw_general-ar}
\end{IEEEeqnarray}

For an even formally self-dual code, according to~\cite[p.~378]{HuffmanPless03_1}, we know that $A_w=A_{n-w}$ for $w$ even and $A_w=0$ for $w$ odd, in~\eqref{eq:weight_enumerator}. Thus, there are at most $\bigl\lfloor\frac{n}{4}\bigr\rfloor+1$ nonzero coefficients $A_w$. For instance, if we want to determine the coefficients of the term corresponding to $A_4$, this would only be possible if we set $j=2$, $h=\ell=0$ or $j=h=0$, $\ell=1$ in~\eqref{eq:Aw_general-ar}, which yields
\begin{IEEEeqnarray*}{rCl}
  A_4 & = &\Biggl(\sum_{r=0}^{\lfloor \nicefrac{n}{8} \rfloor} a_r\biggl({\binom{\nicefrac{n}{2} - 4r}{2}+ \underbrace{7 \binom{2r}{1}}_{14r}}\biggr) \Biggr)x^{n-4}y^4
  \\
  & = & a_0 \binom{\nicefrac{n}{2}}{2} +  a_1 \left(  \binom{\nicefrac{n}{2}-4}{2} +14 \right)
  +a_2\biggl(\binom{\nicefrac{n}{2}-8}{2} 
  \nonumber\\[1mm]
  && \hspace{0.35cm} +\>28 \biggr) + a_3 \biggl(\binom{\nicefrac{n}{2}-12}{2} +42 \biggr)+\cdots.
  \IEEEeqnarraynumspace
\end{IEEEeqnarray*}

For ease of illustration, we compute more terms of~\eqref{eq:Aw_general-ar}: 
\begin{IEEEeqnarray*}{rCl}
  A_0 & = &\sum_{r=0}^{\lfloor\nicefrac{n}{8}\rfloor} a_r,\quad A_2 =\sum_{r=0}^{\lfloor \nicefrac{n}{8} \rfloor} a_r \left(\tfrac{n}{2}-4r\right),
  \nonumber\\[1mm]
  A_6 & = &\sum_{r=0}^{\lfloor \nicefrac{n}{8} \rfloor} a_r \left({\binom{ \nicefrac{n}{2} - 4r}{3}+14r\left(\tfrac{n}{2} - 4r\right)} \right),
  \nonumber\\[1mm]
  A_8 & = & \sum_{r=0}^{\lfloor \nicefrac{n}{8} \rfloor} a_r \Biggl(\binom{\nicefrac{n}{2}-4r}{4}+14r\binom{\nicefrac{n}{2}-4r}{2}\nonumber\\
  &&\qquad\qquad +\>49\binom{2r}{2}-48 r)\Biggr).\IEEEeqnarraynumspace
\end{IEEEeqnarray*}
	

As a result, we can obtain 
the $\bigl\lfloor\frac{n}{8}\bigr\rfloor+1$ unknown coefficients $a_r$, $r\in [0:\bigl\lfloor\frac{n}{8}\bigr\rfloor]$ by solving the system of $\bigl\lfloor\frac{n}{4}\bigr\rfloor+1$ linear equations in~\eqref{eq:Aw_general-ar}. 
The uniqueness of the set of coefficients $a_r$ follows from Gleason's Theorem~\cite[Th.~9.2.1]{HuffmanPless03_1}.

\section{Construction of Isodual Codes from Rate $\nicefrac{1}{2}$ Binary Convolutional codes}
\label{sec:isodual_coco}

An $(n,k)$ binary convolutional code $\code{C}$ is a $k$-dimensional subspace of $\Field_2(D)^{n}$, where $D$ is an indeterminate variable and $\Field_2(D)$ consists of all rational functions in $D$. For a background on convolutional codes, please see, e.g.,~\cite{lincostello04_1}. It is well known~\cite{BocharovaJohannessonKudryashovStahl02_1} that \emph{tailbiting convolutional codes} often produce very competitive linear codes. We point out the following property of the linear block codes obtained by tailbiting applied to convolutional codes of rate 1/2.

\begin{proposition}
  Let $\code{C}$ be a $(2,1)$ binary convolutional code. Then, any $[2k,k ]$ linear code $\code{C}_{\textnormal{tb}}$ obtained from $\code{C}$ by tailbiting is isodual, where $k\geq (m+1)$ and $m$ is the maximum degree of the generator polynomials for $\code{C}$. 
\end{proposition}
\begin{IEEEproof}
  For brevity, we prove this by an example of the convolutional code generated by the \emph{minimal} generator matrix
  \begin{IEEEeqnarray*}{rCl}
    \mat{G}(D) & = &
    \begin{pmatrix}
      g_1(D) & g_2(D)
    \end{pmatrix}
    \nonumber\\[1mm]
    & = &
    \begin{pmatrix}
      a + c D + e D^2 & b + d D + f D^2
    \end{pmatrix}\IEEEeqnarraynumspace
  \end{IEEEeqnarray*}
and its associated $[2\times 5,1\times 5]=[10,5]$ linear code $\code{C}_{\textnormal{tb}}$  by tailbiting for $k=5$. The proof is easily adapted to other tailbiting codes for different code dimensions $k$ and other convolutional codes, but the matrices involved tend to not fit nicely in a page.

It is well known \cite{SolomonVanTilborg79_1},~\cite[p.~107]{MaWolf86_1} that a generator matrix of the linear code $\code{C}_{\textnormal{tb}}$ can be written as 
\begin{equation}
  \mat{G}_{\textnormal{tb}} = \begin{pmatrix}
    a & b & c & d & e & f & & & & \\
    & & a & b & c & d & e & f & & \\
    & & & & a & b & c & d & e & f \\
    e & f & & & & & a & b & c & d \\
    c & d & e & f & & & & & a & b     
  \end{pmatrix},\IEEEeqnarraynumspace \label{eq:genmat_tb}
\end{equation}
and that a parity check matrix for $\code{C}_{\textnormal{tb}}$ can be written as
\begin{IEEEeqnarray*}{c}
    \mat{H}_{\textnormal{tb}} =
  \begin{pmatrix}
    b & a &   &   &   &   & f & e & d & c \\
    d & c & b & a &   &   &  &  & f & e \\
    f & e & d & c & b & a &   &   &  &  \\
    &   & f & e & d & c & b & a &  &      \\
    &   &   &   & f & e & d & c & b & a     
  \end{pmatrix}.
\end{IEEEeqnarray*}

Clearly, for binary codes, $\mat{G}_{\textnormal{tb}}\trans{\mat{H}_{\textnormal{tb}}} = \mat{0}$, and $\mat{G}_{\textnormal{tb}}$ and $ \mat{H}_{\textnormal{tb}}$ generate $[10,5]$ linear codes that are mutually reversed with respect to order of coordinates, and hence they are isodual (thus, they share the same weight enumerator as well).
\end{IEEEproof}

\begin{remark}
  \begin{itemize}
  \item A $[2k,k]$ tailbiting code for any integer  $k \geq (m+1)$, 
    is generated by a matrix constructed like the one in (\ref{eq:genmat_tb}), with the first $k-m$ rows containing successive  two-coordinate shifts of the generator polynomial's coefficients and the last $m$ rows wrapping around like in (\ref{eq:genmat_tb}).
  \item Consider a convolution code $\code{C}$ with 
  free distance $d_\textnormal{free}$. It is well known that the minimum distance $d_{\textnormal{tb}}$ of the tailbiting code $\code{C}_\textnormal{tb}$ is bounded as $d_{\textnormal{tb}}\leq d_{\textnormal{free}}$, 
    and that $d_\textnormal{tb} = d_{\textnormal{free}}$ for any dimension $k \geq k_{\code{C}}$, where $k_{\code{C}}$ is a modest lower threshold that depends only on $\code{C}$. 
  \item The exact weight enumerators, as presented in Appendix~\ref{sec:weight-enumerators-codes_TableI} of this paper, of isodual tailbiting codes, indicated by ``tb'', are conveniently computed by a modified Viterbi algorithm. A straightforward application of this algorithm has a complexity of $O(k \cdot 2^{2m})$.
  \end{itemize}
\end{remark}

}
\ifthenelse{\boolean{short_version}}{\IEEEtriggeratref{6}}{\IEEEtriggeratref{24}}

\bibliographystyle{IEEEtran}
\bibliography{defshort1,biblioHY}

\ifthenelse{\boolean{short_version}}{}{
  \makeatletter\afterpage{\if@firstcolumn \else\afterpage{
      \onecolumn
\begin{landscape}
\section{Weight Enumerators of Codes for Table~\ref{tab:table_secrecy-gains_FU-lattices}}
\label{sec:weight-enumerators-codes_TableI}

\begin{table*}[htbp!]
  \centering
  \caption{Codes and Their Weight Enumerators}
  \vskip -2.0ex
  \Scale[1.0]{
   \begin{IEEEeqnarraybox}[
    \IEEEeqnarraystrutmode
    \IEEEeqnarraystrutsizeadd{3pt}{3pt}]{V/c/V/c/V/c/V/c/V/c/V}
    \IEEEeqnarrayrulerow\\
    & \code{C}
    && \textnormal{Type} 
    && \textnormal{Reference}
    && W_\code{C}(x,y)
    && \chi^\ast_{\Lambda_\textnormal{A}(\code{C})}
    &\\    
    \hline\hline
    & [6,3,2]  && \textnormal{efsd} && \textnormal{\cite{HuffmanPless03_1}} &&  x^6+3x^2y^4+3x^4y^2+y^6  && 1 &\\
    \IEEEeqnarrayrulerow\\
      & [6,3,3]  && \textnormal{ofsd} && \textnormal{\cite{BetsumiyaHarada01_1}}
      && x^6+4x^3y^3+3x^2y^4&& \mathbf{1. 172}&\\
       \IEEEeqnarrayrulerow\\
       & [8,4,4]  && \textnormal{sd} && \textnormal{\cite{HuffmanPless03_1}} && x^8+14x^4y^4+y^8 && \mathbf{1. 333} & \\
       \IEEEeqnarrayrulerow\\
        & [8,4,3]  && \textnormal{ofsd} && \textnormal{\cite{BetsumiyaHarada01_1}} && x^8+3x^5y^3+7x^4y^4+4x^3y^5+xy^7 && 1. 282& \\
       \IEEEeqnarrayrulerow\\
      & [8,4,3]  && \textnormal{ofsd} && \textnormal{\cite{BetsumiyaHarada01_1}} && x^8+4x^5y^3+5x^4y^4+4x^3y^5+2x^2y^6 && 1. 264& \\
       \IEEEeqnarrayrulerow\\ 
        & [10,5,4]  && \textnormal{efsd} && \textnormal{\cite{HuffmanPless03_1}} && x^{10}+15x^6y^4+15x^4y^6+y^{10} && 1. 455& \\
       \IEEEeqnarrayrulerow\\ 
          & [10,5,4]  && \textnormal{ofsd} && \textnormal{\cite{BetsumiyaHarada01_1}} && x^{10}+10x^6y^4+16x^5y^5+5x^2y^8 && \mathbf{1.478} & \\
       \IEEEeqnarrayrulerow\\ 
          & [12,6,4]  && \textnormal{sd} && \textnormal{\cite{LinOggier13_1}} && x^{12}+15x^8y^4+32x^6y^6+15x^4y^8+y^{12} && 1. 6 & \\
       \IEEEeqnarrayrulerow\\ 
       & [12,6,4]  && \textnormal{efsd} && \textnormal{\cite{BetsumiyaGulliverHarada99_1}} && x^{12}+15x^8y^4+32x^6y^6+15x^4y^8+y^{12} && 1. 6 & \\
       \IEEEeqnarrayrulerow\\ 
        & [12,6,4]  && \textnormal{ofsd} && \textnormal{\cite{BetsumiyaHarada01_1}} && x^{12}+6x^8y^4+24x^7y^5+16x^6y^6+9x^4y^8+8x^3y^9 && \mathbf{1. 657} & \\
       \IEEEeqnarrayrulerow\\ 
         & [14,7,4]  && \textnormal{sd} && \textnormal{\cite{LinOggier13_1}} && x^{14}+14x^{10}y^4+49x^8y^6+49x^6y^8+14x^4y^{10}+y^{14} && 1.778 & \\
       \IEEEeqnarrayrulerow\\ 
       & [14,7,2]  && \textnormal{efsd} && \textnormal{\cite{BetsumiyaGulliverHarada99_1}} && x^{14}+x^{12} y^2+15 x^{10} y^4+47 x^8 y^6+47 x^6 y^8+15 x^4 y^{10}+x^2 y^{12}+y^{14} && 1.6 & \\
       \IEEEeqnarrayrulerow\\ 
         & [14,7,4]  && \textnormal{ofsd} && \textnormal{\cite{BetsumiyaHarada01_1}} && x^{14}+3x^{10}y^4+24x^9y^5+36x^8y^{6}+16x^7y^7 + 11x^6y^8 + 24x^5y^9+12x^4y^{10}+x^2y^{12} && \mathbf{1.875} & \\
       \IEEEeqnarrayrulerow\\ 
       & [16,8,4]  && \textnormal{sd} && \textnormal{\cite{LinOggier13_1}} && x^{16}+12x^{12}y^4+64x^{10}y^6+102x^8y^8+64x^6y^{10}+12x^4y^{12}+y^{16} && 2 & \\
       \IEEEeqnarrayrulerow\\ 
       & [16,8,4]  && \textnormal{efsd} && \textnormal{\cite{BetsumiyaHarada01_2}} && x^{16}+ 4 x^{12} y^4+96 x^{10} y^6+54 x^8 y^8+96 x^6 y^{10}+4 x^4 y^{12}+x^{16}+y^{16} && 2.133 & \\
       \IEEEeqnarrayrulerow\\ 
     & [16,8,5]  && \textnormal{ofsd} && \textnormal{\cite{BetsumiyaHarada01_1}} && x^{16}+24x^{11}y^5+44x^{10}y^6+40x^9y^7+45x^8y^{8}+40x^7y^9+28x^6y^{10}+24x^5y^{11}+10x^4y^{12} && \mathbf{2.141} & \\
       \IEEEeqnarrayrulerow\\ 
       & [18,9, 4]  && \textnormal{sd} && \textnormal{\cite{LinOggier13_1}} && x^{18}+9x^{14}y^4+75x^{12}y^6+171x^{10}y^8+171x^{8}y^{10}+75x^{6}y^{12}+9x^4y^{14}+y^{18} &&2.286 & \\
       \IEEEeqnarrayrulerow\\ 
        & [18,9, 6]  && \textnormal{efsd} && \textnormal{\cite{SloaneHeninger06_1}} &&  x^{18}+102x^{12}y^6+153x^{10}y^8+153x^{8}y^{10}+102x^{6}y^{12}+y^{18} && \mathbf{2.485} & \\
       \IEEEeqnarrayrulerow\\ 
       & [18,9, 5]  && \textnormal{ofsd} &&
       \textnormal{tb}
       && {\tiny x^{18}+18x^{13}y^5+48x^{12}y^6+63x^{11}y^7+81x^{10}y^{8}+100x^{9}y^{9}+72x^8y^{10}+54x^7y^{11}+54x^6y^{12}+18x^5y^{13}+3x^{3}y^{15}} && 2.424 & \\
     \IEEEeqnarrayrulerow \\
     & [20,10,4]  && \textnormal{sd} && \textnormal{\cite{LinOggier13_1}} && x^{20}+5x^{18}y^4+80x^{14}y^6+250x^{12}y^{8}+352x^{10}y^{10}+250x^{8}y^{12}+80x^{6}y^{14}+5x^{4}y^{18}+y^{20} && 2.523 &\\
    \IEEEeqnarrayrulerow\\
     & [20,10,6]  && \textnormal{efsd} && \textnormal{\cite{FieldsGaboritHuffmanPless01_1}} && x^{20}+90x^{14}y^6+255x^{12}y^8+332x^{10}y^{10}+255x^{8}y^{12}+90x^{6}y^{14}+y^{20} && 2.813 &\\
    \IEEEeqnarrayrulerow\\
     & [20,10,6]  && \textnormal{ofsd} &&  \textnormal{\cite{BetsumiyaGulliverHarada99_1}}   && x^{20}+40 x^{14} y^6+160 x^{13} y^7+130 x^{12} y^8+176 x^{10} y^{10}+320 x^9 y^{11}+120 x^8 y^{12}+40 x^6 y^{14}+32 x^5 y^{15}+5 x^4 y^{16} && \mathbf{2.868} &\\
    \IEEEeqnarrayrulerow
  \end{IEEEeqnarraybox}}
\end{table*}
\end{landscape}


\begin{landscape}
\begin{table*}[htbp!]
  \centering
  \caption{Codes and Their Weight Enumerators - Cont.}
   \begin{IEEEeqnarraybox}[
    \IEEEeqnarraystrutmode
    \IEEEeqnarraystrutsizeadd{3pt}{3pt}]{V/c/V/c/V/c/V/c/V/c/V}
    \IEEEeqnarrayrulerow\\
    & \code{C}
    && \textnormal{Type} 
    && \textnormal{Reference}
    && W_\code{C}(x,y)
    && \chi^\ast_{\Lambda_\textnormal{A}(\code{C})}
    &\\    
    \hline\hline
     & [22,11,6]  && \textnormal{sd} && \textnormal{\cite{LinOggier13_1}} && x^{22}+77x^{16}y^6+330x^{14}y^8+616x^{12}y^{10}+616x^{10}y^{12}+330x^{8}y^{14}+77x^{6}y^{16}+y^{22} && 3.2 &\\
    \IEEEeqnarrayrulerow\\
    & [22,11,6]  && \textnormal{ofsd} &&
    \textnormal{tb}
    && \pbox{16cm}{\centering \vspace{0.1cm}
      $x^{22}+44x^{16}y^6+121x^{15}y^7+143x^{14}y^{8}+231x^{13}y^{9}+319x^{12}y^{10}+298x^{11}y^{11}+330x^{10}y^{12}+286x^{9}y^{13}+154x^{8}y^{14}+77x^{7}y^{15}+22x^{6}y^{16}+11x^{5}y^{17}+11x^{4}y^{18}$ \vspace{0.1cm}
    } && 3.243 & \\
    \IEEEeqnarrayrulerow\\
     & [22,11,7]  && \textnormal{ofsd} && \textnormal{\cite{BetsumiyaHarada01_1}}
     && x^{22}+176x^{15}y^7+330x^{14}y^8+672x^{11}y^{11}+616x^{10}y^{12}+176x^{7}y^{15}+77x^{6}y^{16} && \mathbf{3.335} &\\
    \IEEEeqnarrayrulerow\\
     & [24,12,8]  && \textnormal{sd} && \textnormal{\cite{MacWilliamsSloane77_1}} && x^{24} + 759 x^{16} y^{8} + 2576 x^{12} y^{12} + 759 x^{8} y^{16} +  y^{24} && \mathbf{3.879} &\\
    \IEEEeqnarrayrulerow\\
     & [24,12,6]  && \textnormal{efsd} &&
     \textnormal{tb}
     && x^{24}+ 64 x^{18} y^6+375 x^{16} y^8+960 x^{14} y^{10}+1296 x^{12} y^{12}+960 x^{10} y^{14}+375 x^8 y^{16}+64 x^6 y^{18}+y^{24} && 3.657 &\\
    \IEEEeqnarrayrulerow\\
       & [30,15,6]  && \textnormal{sd} && \textnormal{\cite{ConwaySloane90_1}} &&\pbox{16cm}{\centering \vspace{0.1cm} $ x^{30}+19 x^{24} y^6+393 x^{22} y^8+1848 x^{20} y^{10}+5192 x^{18} y^{12}+8391 x^{16} y^{14}+8391 x^{14} y^{16}+5192 x^{12} y^{18}+1848 x^{10} y^{20}+393 x^8 y^{22}+19 x^6 y^{24}+y^{30}$\vspace{0.1cm}} && 5.697 &\\
     \IEEEeqnarrayrulerow\\
     & [30,15,8]  && \textnormal{efsd} && \textnormal{\cite{BouyuklievaBouyukliev10_1}} && x^{30}+ 450x^{22}y^8 + 1848x^{20}y^{10}+5040x^{18}y^{12}+9045x^{16}y^{14}+9045x^{14}y^{16}+5040x^{12}y^{18}+1848x^{10}y^{20}+450x^{8}y^{22}+y^{30}&& \mathbf{5.843} &\\   
    \IEEEeqnarrayrulerow\\
     & [30,15,7]  && \textnormal{ofsd} &&
     \textnormal{tb}
     && \pbox{16cm}{\centering \vspace{0.1cm} $x^{30} + 60 x^{23} y^7+210 x^{22} y^8+500 x^{21} y^9+930 x^{20} y^{10}+1560 x^{19} y^{11}+2570 x^{18} y^{12}+3660 x^{17} y^{13}+4530 x^{16} y^{14}+4824 x^{15} y^{15}+4335 x^{14} y^{16}+3660 x^{13} y^{17}+2710 x^{12} y^{18}+1560 x^{11} y^{19}+918 x^{10} y^{20}+500 x^9 y^{21}+150 x^8 y^{22}+60 x^7 y^{23}+30 x^6 y^{24}$ \vspace{0.1cm}} && 5.785 &\\  
    \IEEEeqnarrayrulerow\\
    & [32,16,8]  && \textnormal{sd} && \textnormal{\cite{SloaneHeninger06_1}} && x^{32}+ 620 x^{24} y^8+13888 x^{20} y^{12}+36518 x^{16} y^{16}+13888 x^{12} y^{20}+620 x^8 y^{24}+y^{32} && 6.564 &\\
    \IEEEeqnarrayrulerow \\
    & [32,16,8]  && \textnormal{sd} && \textnormal{\cite{SloaneHeninger06_1}} && \pbox{16cm}{\centering \vspace{0.1cm} $x^{32}+364 x^{24} y^8+2048 x^{22} y^{10}+6720 x^{20} y^{12}+14336 x^{18} y^{14}+18598 x^{16} y^{16}+14336 x^{14} y^{18}+6720 x^{12} y^{20}+2048 x^{10} y^{22}+364 x^8 y^{24}+y^{32} \vspace{0.1cm}$} && 6.737 &\\
    \IEEEeqnarrayrulerow \\
    & [32,16,8]  && \textnormal{efsd} &&
    \textnormal{tb}
    && \pbox{16cm}{\centering \vspace{0.1cm} $x^{32}+348 x^{24} y^8+2176 x^{22} y^{10}+6272 x^{20} y^{12}+15232 x^{18} y^{14}+17478 x^{16} y^{16}+15232 x^{14} y^{18}+6272 x^{12} y^{20}+2176 x^{10} y^{22}+348 x^8 y^{24}+y^{32}$\vspace{0.1cm}} && \mathbf{6.748} &\\
    \IEEEeqnarrayrulerow \\
    & [32,16,7]  && \textnormal{ofsd} &&
    \textnormal{tb}
    && \pbox{16cm}{\centering \vspace{0.1cm} $x^{32}+64 x^{25} y^7+176 x^{24} y^8+384 x^{23} y^9+984 x^{22} y^{10}+2096 x^{21} y^{11}+3500 x^{20} y^{12}+5136 x^{19} y^{13}+7096 x^{18} y^{14}+8624 x^{17} y^{15}+9133 x^{16} y^{16}+8848 x^{15} y^{17}+7384 x^{14} y^{18}+5136 x^{13} y^{19}+3292 x^{12} y^{20}+1968 x^{11} y^{21}+1032 x^{10} y^{22}+464 x^9 y^{23}+154 x^8 y^{24}+48 x^7 y^{25}+16 x^6 y^{26}$\vspace{0.1cm}} && 6.628 &\\
    \IEEEeqnarrayrulerow \\
     & [40,20,8]  && \textnormal{sd} && \textnormal{\cite{ConwaySloane90_1}} && x^{40}+285 x^{32} y^8+21280 x^{28} y^{12}+239970 x^{24} y^{16}+525504 x^{20} y^{20}+239970 x^{16} y^{24}+21280 x^{12} y^{28}+285 x^8 y^{32}+y^{40}&& 11.977 &\\
    \IEEEeqnarrayrulerow \\
    & [40,20,8]  && \textnormal{sd} && \textnormal{\cite{ConwaySloane90_1}} && \pbox{16cm}{\centering \vspace{0.1cm} $x^{40}+125 x^{32} y^8+1664 x^{30} y^{10}+10720 x^{28} y^{12}+44160 x^{26} y^{14}+119810 x^{24} y^{16}+216320 x^{22} y^{18}+262976 x^{20} y^{20}+216320 x^{18} y^{22}+119810 x^{16} y^{24}+44160 x^{14} y^{26}+10720 x^{12} y^{28}+1664 x^{10} y^{30}+125 x^8 y^{32}+y^{40}$ \vspace{0.1cm}} && 12.191 &\\
    \IEEEeqnarrayrulerow \\
    & [40,20,8]  && \textnormal{efsd} &&
    \textnormal{tb}
    && \pbox{16cm}{\centering \vspace{0.1cm} $x^{40}+150 x^{32} y^8+1564 x^{30} y^{10}+10770 x^{28} y^{12}+44460 x^{26} y^{14}+119385 x^{24} y^{16}+216120 x^{22} y^{18}+263676 x^{20} y^{20}+216120 x^{18} y^{22}+119385 x^{16} y^{24}+44460 x^{14} y^{26}+10770 x^{12} y^{28}+1564 x^{10} y^{30}+150 x^8 y^{32}+y^{40}$ \vspace{0.1cm}} && 12.134 &\\
    \IEEEeqnarrayrulerow \\
    & [40,20,9]  && \textnormal{ofsd} &&
    \textnormal{tb}
    && \pbox{16cm}{\centering \vspace{0.1cm} $x^{40}+360 x^{31} y^9+922 x^{30} y^{10}+2060 x^{29} y^{11}+5775 x^{28} y^{12}+11340 x^{27} y^{13}+20980 x^{26} y^{14}+39064 x^{25} y^{15}+60185 x^{24} y^{16}+83680 x^{23} y^{17}+109740 x^{22} y^{18}+125640 x^{21} y^{19}+130046 x^{20} y^{20}+125640 x^{19} y^{21}+107680 x^{18} y^{22}+83680 x^{17} y^{23}+60830 x^{16} y^{24}+39064 x^{15} y^{25}+22250 x^{14} y^{26}+11340 x^{13} y^{27}+4755 x^{12} y^{28}+2060 x^{11} y^{29}+1084 x^{10} y^{30}+360 x^9 y^{31}+40 x^8 y^{32}$\vspace{0.1cm} } && \mathbf{12.364} &\\
    \IEEEeqnarrayrulerow  \\
     & [42,21,10]  && \textnormal{efsd} &&
     \textnormal{tb}
     && \pbox{16cm}{\centering \vspace{0.1cm} $x^{42}+1722 x^{32} y^{10}+10619 x^{30} y^{12}+49815 x^{28} y^{14}+157563 x^{26} y^{16}+341530 x^{24} y^{18}+487326 x^{22} y^{20}+487326 x^{20} y^{22}+341530 x^{18} y^{24}+157563 x^{16} y^{26}+49815 x^{14} y^{28}+10619 x^{12} y^{30}+1722 x^{10} y^{32}+y^{42}$ \vspace{0.1cm} } && 14.482 &\\
    \IEEEeqnarrayrulerow 
  \end{IEEEeqnarraybox}
\end{table*}
\end{landscape}


\begin{landscape}
\begin{table*}[htbp!]
  \centering
  \caption{Codes and Their Weight Enumerators - Cont.}
  \vskip -2.0ex
   \begin{IEEEeqnarraybox}[
    \IEEEeqnarraystrutmode
        \IEEEeqnarraystrutsizeadd{7pt}{7pt}]{V/c/V/c/V/c/V/c/V/c/V}
    \IEEEeqnarrayrulerow\\
    & \code{C}
    && \textnormal{Type} 
    && \textnormal{Reference}
    && W_\code{C}(x,y)
    && \chi^\ast_{\Lambda_\textnormal{A}(\code{C})}
    &\\    
    \hline\hline
    & [56,28,12]  && \textnormal{{efsd}} &&
    \textnormal{tb}
    && \pbox{16cm}{\centering \vspace{0.1cm} $x^{56}+4634 x^{44} y^{12}+44828 x^{42} y^{14}+307650 x^{40} y^{16}+1575924 x^{38} y^{18}+5865384 x^{36} y^{20}+15969660 x^{34} y^{22}+32430013 x^{32} y^{24}+ 49502068 x^{30} y^{26}+57035132 x^{28} y^{28}+ 49502068 x^{26} y^{30}+32430013 x^{24} y^{32}+15969660 x^{22} y^{34}+5865384 x^{20} y^{36}+1575924 x^{18} y^{38}+307650 x^{16} y^{40}+44828 x^{14} y^{42}+ 4634 x^{12} y^{44}+y^{56}$\vspace{0.1cm} }&& 42.838 & \\
    \IEEEeqnarrayrulerow\\
     & [70,35,12]  && \textnormal{sd} && \textnormal{\cite{Harada97_1}} &&  \pbox{16cm}{\centering \vspace{0.1cm} $x^{70}+ 832 x^{58} y^{12}+10770 x^{56} y^{14}+142279 x^{54} y^{16}+1353320 x^{52} y^{18}+9437352 x^{50} y^{20}+49957193 x^{48} y^{22}+204165154 x^{46} y^{24}+650426976 x^{44} y^{26}+1627816992 x^{42} y^{28}+3221537516 x^{40} y^{30}+5066102223 x^{38} y^{32}+6348918576 x^{36} y^{34}+6348918576 x^{34} y^{36}+5066102223 x^{32} y^{38}+3221537516 x^{30} y^{40}+1627816992 x^{28} y^{42}+650426976 x^{26} y^{44}+204165154 x^{24} y^{46}+49957193 x^{22} y^{48}+9437352 x^{20} y^{50}+1353320 x^{18} y^{52}+142279 x^{16} y^{54}+10770 x^{14} y^{56}+832 x^{12} y^{58}+y^{70}$ \vspace{0.1cm}} && 127.712 &\\
     \IEEEeqnarrayrulerow \\
     & [70,35,12]  && \textnormal{efsd} && 
     \textnormal{tb}
     &&  \pbox{16cm}{\centering \vspace{0.1cm} $x^{70}  + 455  x^{58}  y^{12} + 11235  x^{56}  y^{14} + 145985  x^{54}  y^{16} + 1348130  x^{52}  y^{18} + 9430974  x^{50}  y^{20} + 49926695  x^{48}  y^{22} + 204318835  x^{46}  y^{24} + 650297655  x^{44}  y^{26} + 1627628010  x^{42}  y^{28} + 3221888194  x^{40}  y^{30} + 5066010495  x^{38}  y^{32} + 6348862520  x^{36}  y^{34} + 6348862520  x^{34}  y^{36} + 5066010495  x^{32}  y^{38} + 3221888194  x^{30}  y^{40} + 1627628010  x^{28}  y^{42} + 650297655  x^{26}  y^{44} + 204318835  x^{24}  y^{46} + 49926695  x^{22}  y^{48} + 9430974  x^{20}  y^{50} + 1348130  x^{18}  y^{52} + 145985  x^{16}  y^{54} + 11235  x^{14}  y^{56} + 455  x^{12}  y^{58} +  y^{70}$ \vspace{0.1cm}} && 128.073 &\\
     \IEEEeqnarrayrulerow \\
    & [70,35,13] && \textnormal{ofsd} &&
    \textnormal{tb}
    && \pbox{16cm}{\centering \vspace{0.1cm} $x^{70}+1225 x^{57} y^{13}+6125 x^{56} y^{14}+21700 x^{55} y^{15}+72590 x^{54} y^{16}+232680 x^{53} y^{17}+676410 x^{52} y^{18}+1838375 x^{51} y^{19}+4711427 x^{50} y^{20}+11204975 x^{49} y^{21}+24964310 x^{48} y^{22}+52191335 x^{47} y^{23}+102128145 x^{46} y^{24}+187879531 x^{45} y^{25}+325261230 x^{44} y^{26}+529884495 x^{43} y^{27}+813742900 x^{42} y^{28}+1178595250 x^{41} y^{29}+1610725606 x^{40} y^{30}+2078727420 x^{39} y^{31}+2533396005 x^{38} y^{32}+2916830420 x^{37} y^{33}+3174375820 x^{36} y^{34}+3264970134 x^{35} y^{35}+3174028690 x^{34} y^{36}+2917093830 x^{33} y^{37}+2533383720 x^{32} y^{38}+2078410810 x^{31} y^{39}+1610915418 x^{30} y^{40}+1178784530 x^{29} y^{41}+813674900 x^{28} y^{42}+529809070 x^{27} y^{43}+325223220 x^{26} y^{44}+187929077 x^{25} y^{45}+102154885 x^{24} y^{46}+52153640 x^{23} y^{47}+24962700 x^{22} y^{48}+11215020 x^{21} y^{49}+4706842 x^{20} y^{50}+1841315 x^{19} y^{51}+682115 x^{18} y^{52}+232155 x^{17} y^{53}+69930 x^{16} y^{54}+20727 x^{15} y^{55}+5845 x^{14} y^{56}+1435 x^{13} y^{57}+350 x^{12} y^{58}+35 x^{11} y^{59}$ \vspace{0.1cm}} && \mathbf{128.368} & \\
    \IEEEeqnarrayrulerow \\
    & [78,39,14] && \textnormal{{efsd}} &&
    \textnormal{tb}
    && \pbox{16cm}{\centering \vspace{0.1cm} $x^{78}+3471 x^{64} y^{14}+63336 x^{62} y^{16}+772980 x^{60} y^{18}+7219368 x^{58} y^{20}+51527346 x^{56} y^{22}+287551706 x^{54} y^{24}+1266693912 x^{52} y^{26}+4442835540 x^{50} y^{28}+12510913844 x^{48} y^{30}+28453167444 x^{46} y^{32}+52493946648 x^{44} y^{34}+78823802720 x^{42} y^{36}+96539408628 x^{40} y^{38}+96539408628 x^{38} y^{40}+78823802720 x^{36} y^{42}+52493946648 x^{34} y^{44}+28453167444 x^{32} y^{46}+12510913844 x^{30} y^{48}+4442835540 x^{28} y^{50}+1266693912 x^{26} y^{52}+287551706 x^{24} y^{54}+51527346 x^{22} y^{56}+7219368 x^{20} y^{58}+772980 x^{18} y^{60}+63336 x^{16} y^{62}+3471 x^{14} y^{64}+y^{78}$ \vspace{0.1cm}} && 241.042 & \\
    \IEEEeqnarrayrulerow \\
    & [108, 54, 14] && \textnormal{efsd} && 
    \textnormal{tb}
    && \pbox{16cm}{\centering \vspace{0.1cm} $ x^{108}+756 x^{94} y^{14}+5022 x^{92} y^{16}+30354 x^{90} y^{18}+371223 x^{88} y^{20}+5418846 x^{86} y^{22}+71085987 x^{84} y^{24}+765738684 x^{82} y^{26}+6738702390 x^{80} y^{28}+48969093384 x^{78} y^{30}+296438923962 x^{76} y^{32}+1505875815558 x^{74} y^{34}+6456109668648 x^{72} y^{36}+23473804361040 x^{70} y^{38}+72678688668432 x^{68} y^{40}+192289983824466 x^{66} y^{42}+436005471914253 x^{64} y^{44}+849263560631748 x^{62} y^{46}+1423721807648100 x^{60} y^{48}+2057133110131674 x^{58} y^{50}+2564434300382478 x^{56} y^{52}+2759767104647972 x^{54} y^{54}+2564434300382478 x^{52} y^{56}+2057133110131674 x^{50} y^{58}+1423721807648100 x^{48} y^{60}+849263560631748 x^{46} y^{62}+436005471914253 x^{44} y^{64}+192289983824466 x^{42} y^{66}+72678688668432 x^{40} y^{68}+23473804361040 x^{38} y^{70}+6456109668648 x^{36} y^{72}+1505875815558 x^{34} y^{74}+296438923962 x^{32} y^{76}+48969093384 x^{30} y^{78}+6738702390 x^{28} y^{80}+765738684 x^{26} y^{82}+71085987 x^{24} y^{84}+5418846 x^{22} y^{86}+371223 x^{20} y^{88}+30354 x^{18} y^{90}+5022 x^{16} y^{92}+756 x^{14} y^{94}+y^{108}$ \vspace{0.1cm}} && 2573.53 & \\
     \IEEEeqnarrayrulerow  
  \end{IEEEeqnarraybox}    
\end{table*}
\end{landscape}
    }
    \fi}\makeatother
}

\end{document}